\documentclass[useAMS,usenatbib]{mn2e}

\title[GeV gamma-rays and TeV neutrinos from binary systems]
{GeV gamma-rays and TeV neutrinos from very massive compact binary systems:
The case of WR 20a}
\author[W. Bednarek]{W. Bednarek\thanks{E-mail:
bednar@fizwe4.fic.uni.lodz.pl}\\
Department of experimental Physics, University of \L \'od\'z,
ul. Pomorska 149/153, 90-236 \L \'od\'z, Poland}
\begin{document}

\date{Accepted . Received ; in original form }

\pagerange{\pageref{firstpage}--\pageref{lastpage}} \pubyear{2002}

\maketitle

\label{firstpage}

\begin{abstract}
Massive Wolf-Rayet stars in a compact binary system are characterized by 
very strong winds which collide creating a shock wave. If the wind nuclei, 
from  
helium up to oxygen, accelerated  at the shock can reach large enough energies, 
they suffer disintegration in collisions with soft thermal radiation from the massive 
stars injecting relativistic protons and neutrons. 
Protons collide with the matter of the wind 
and a fraction of neutrons collide with the massive stars producing $\gamma$-rays and 
neutrinos via decay of pions. 
We calculate $\gamma$-ray fluxes from inverse Compton $e^\pm$ pair cascades initiated 
by primary $\gamma$-rays 
and leptons, which are produced by protons, and the neutrino fluxes, produced by protons and 
neutrons, for the example compact massive binary WR 20a. 
From normalization of these cascade $\gamma$-ray spectra to the fluxes of the EGRET 
sources, 2EG J1021-5835 and 2EG J1049-5847 observed in the direction of WR 20a,
we conclude that this massive binary can be detected by the IceCube type neutrino
detector. The estimated muon neutrino event rate inside the 1 km$^2$ detector during
1 yr is between a few up to a few tens, depending on the spectrum of accelerated 
nuclei.
\end{abstract}


The observation of neutrinos from discrete sources should give a unique
information on the acceleration of hadrons in our Galaxy and on the origin of cosmic
rays at energies above $\sim 10^{12}$ eV. 
Possible production of neutrinos by relativistic hadrons accelerated inside the binary 
systems has been discussed in the case of accreting neutron stars 
(Gaisser \& Stanev~1985, Kolb, Turner \& Walker~1985, 
Berezinsky, Castagnoli \& Galeotti~1985, 
Anchordoqui et al.~2003), energetic pulsars (Harding \& Gaisser~1990), 
and accreting solar mass black holes called 
microquasars (Levinson \& Waxman~2001, Distefano et al.~2002, Romero et al.~2003).
For recent review of these models see e.g. Bednarek, Burgio \& Montaruli~(2005). 
Most of these models predict neutrino
fluxes which should be easily detectable by the 1 km$^2$ neutrino detector of the 
IceCube type and  in some cases by detectors of the present AMANDA II type. 
Recent searches of the discrete neutrino sources by the AMANDA II detector,
which base on the analysis of the three-year experimental data 
(Ahrens et al.~2004, Ackermann et al. 2005), show
no clear evidence of a significant neutrino flux excess above the background. 

In this paper we show that also compact binary systems containing two massive stars
can be interesting sources of TeV neutrinos likely to be observed by large scale 
neutrino telescopes. In fact, 
between 227 Wolf-Rayet (WR) stars (van der Hucht~2001), 39$\%$ form binary systems 
(including probable binaries), and 
about $15\%$ of them form compact massive binaries with orbital periods 
$<10$ days (see Table 15 in van der Hucht~2001). One extreme example of a binary
system, WR 20a, contains two WR 
stars with masses $\sim 70M_\odot$, surface temperature 
$T = 10^5T_5$ K $\approx 4\times 10^4$ K, and radii $\sim 20R_\odot$ 
(where $R_\odot = 7\times 10^{10}$ cm is the radius of the Sun), on an orbit with 
the semimajor axis $a/2\sim 26R_\odot$ (Rauw et al.~2004, Bonanos et al.~2004).
Let's consider such a very massive binary composed of two massive WR type stars
characterized by very strong winds, with  mass loss rate 
$\dot{M}=10^{-5}\dot{M}_{-5}$ M$_\odot$ yr$^{-1}$$\sim (0.8-8)\times 10^{-5} M_\odot$ 
yr$^{-1}$, wind terminal velocities 
$v_{\rm w}^\infty = 3\times 10^3v_3$ km s$^{-1}$$\sim 
(1-5)\times 10^3$ km s$^{-1}$. 
The radiatively driven stellar
winds accelerate from the surface of the star according to 
$v_3(r) = v_3^0 + (v_3^\infty -  v_3^0)(1 - 1/r)^\beta\approx v_3^\infty (1 - 1/r)$, 
where 
$v_3(r)$ is the wind velocity at the distance, $R = rR_{\rm WR}$, from the star, 
$v_3^0\cong 0.01 v_3^\infty$, $\beta\cong 1$ (e.g. Castor \& Lamers~1979). The 
surface magnetic fields of WR stars can grow up to $B_{\rm s}\sim 10^4$ Gs 
(Maheswaran \& Cassinelli~1994). For the massive stars in WR 20a binary system we
apply the typical values of $B = 10^3B_3$ G, i.e. significantly below $10^4$ G.
Such stellar winds collide creating a double shock structure separated by the contact 
discontinuity which distance from the stars depends on their wind parameters.

It is likely that the binary system WR 20a is responsible for one of the 
EGRET sources observed in this region, i.e. 2EG J1021-5835 and GeV J1025-5809
or 2EG J1049-5847 and GeV J1047-5840 (see Thompson et al.~1995, Lamb \& MaComb~1997). 
They have flat spectra with the index close to 2 (Merck et al.~1996) and fluxes at 
the level of $\sim 10^{-7}$ cm$^{-2}$ s$^{-1}$ above 1 GeV (Lamb \& MaComb~1997).

\begin{figure}
\vskip 4.5truecm
\includegraphics{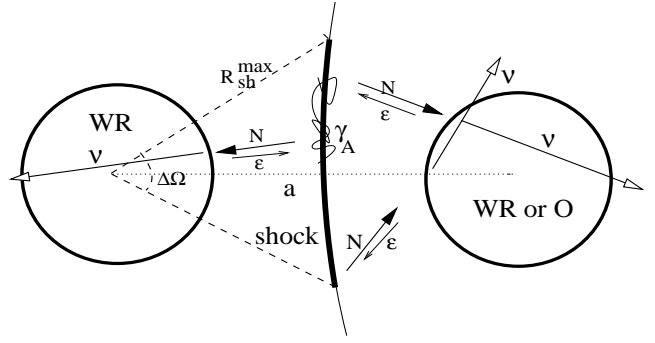}
\caption{Schematic picture of a compact binary of two massive stars 
(WR+WR or WR+O) which create a strong shock as a result of their wind collisions.
Nuclei are accelerated at the shock in the reconnection regions or by the diffusive
shock acceleration mechanism to the Lorentz factors $\gamma_{\rm max}$. 
Only nuclei accelerated within a part of the shock 
(marked by thick line within the solid angle $\Delta\Omega$) are accelerated 
to the Lorentz factors above $\gamma_{\rm min}$ which allow their efficient
photo-disintegration in collisions with thermal photons from the massive stars.
Significant fraction of neutrons dissolved from nuclei move toward the massive stars 
(due to the 
head on collisions) and interact with the matter of the stellar atmospheres producing 
neutrinos.}
\label{fig1}
\end{figure}

Nuclei present in the stellar winds can be accelerated at such shock structure either 
due to the magnetic field reconnection (we call this possibility as model I) or the 
diffusive shock acceleration mechanism (model II) (see Fig.~\ref{fig1}). 
The acceleration occurs in the fast reconnection mode provided that following 
condition is fulfilled, $\beta = 2m_{\rm e}/Am_{\rm p}\approx 10^{-3}/A < 
8\pi \rho k_{\rm B}T_{\rm sh}/B_{\rm sh}^2\approx  
3.3\times 10^{-3}\dot{M}_{-5}/B_3^2r^2v_3(r)$
(although the fast reconnection can also occur in specific situations when this
condition is not met, e.g. Hanasz \& Lesch~2003), assuming typical temperature of the 
plasma at the shock $T_{\rm sh} = 10^7$ K (e.g. Stephens et al.~1992), 
density of the wind 
$\rho = \dot{M}/4\pi R^2v_{\rm w}\approx 5.4\times 
10^{10}\dot{M}_{-5}/r^2v_3(r)$ cm$^{-3}$. 
Then, the reconnection occurs with the speed close to the Alfven velocity 
$v_{\rm A} = B_{\rm sh}/\sqrt{4\pi m_{\rm p}\rho}\approx 
7\times 10^8B_3v_3^{1/2}(r)/r\dot{M}_{-5}^{1/2}$ cm s$^{-1}$. 
Nuclei can reach maximum Lorentz factors of the order of
\begin{eqnarray}
\gamma_{\rm max}\approx {{Ze B_{sh} L_{rec}v_{A}}\over{A m_{\rm p} c}}\rm 
\approx 2.6\times 10^6{{B_3^2 L_{12}v_3^{1/2}(r)}\over{r^3\dot{M}_{-5}^{1/2}}},
\label{eq1}
\end{eqnarray}
where $A/Z=2$, $m_{\rm p}\approx m_{\rm n}$ is the nucleon mass, 
$L_{\rm rec} = 10^{12}L_{12}$ cm is the length of the reconnection region,
$c$ is the velocity of light, and $e$ is the proton charge. In order to estimate 
the magnetic field strength after the shock, we apply the model for external 
magnetic field structure in the case of strong outflowing gas (Usov \& Melrose~1992). 
According to this model the magnetic field is a dipole type below the Alfven radius,
$R_{\rm A}$, becomes radial above $R_{\rm A}$, and at the largest distances, 
determined by the rotation velocity of the star, i.e. at above 
$\sim 10 R_{\rm WR}$,
the toroidal component dominates. We are interested mainly in the region in which the
dipole and radial components dominate.
Above $\sim 10R_{WR}$ the product of the magnetic field strength at the 
shock and the length of the reconnection region is roughly constant and the 
maximum energy of accelerated nuclei does not depend on the distance from the star. 
The Alfven radius for the example parameters considered in this paper,
$v_3^\infty = 1$, $B_3 = 1$, and $\dot{M}_{-5} = 3$, is at
$R_{\rm A}\approx 1.3 R_{\rm WR}$. Therefore, we estimate the magnetic field 
at the shock in the region of radial magnetic field from 
\begin{eqnarray}
B_{\rm sh}(r)\approx B_{\rm s} (R_{\rm WR}/R_{\rm A})^3
(R_{\rm A}/R_{\rm sh})^2\approx 750B_3/r^2~~{\rm Gs},
\label{eq2}
\end{eqnarray}
where $R_{\rm sh}$ 
is the distance from the center of the star to the acceleration region at the shock, 
i.e. $R_{\rm sh} = r R_{\rm WR}$. 

Nuclei can be also accelerated diffusively by the I order Fermi shock
acceleration mechanism. In this case they obtain a power law spectrum. 
Nuclei are accelerated to 
$\gamma_{\rm A}$ during the time estimated by $\tau_{\rm acc}\approx  
(R_{\rm L}/c)(c/v_{\rm w})^2$, where $R_{\rm L} = 
Am_{\rm p}\gamma_{\rm A}/(eZB_{\rm sh})\approx 
8\times 10^9\gamma_6r^2/B_3~~{\rm cm}$ is the Larmor radius required
to complete acceleration to $\gamma_{\rm A} = 10^6\gamma_6$.
The maximum energy of nuclei is not limited by any
radiation loss process but by their escape from the acceleration region. 
We identify this escape mechanism with the convection of relativistic nuclei
with the wind flow along the surface of the shock. The escape time is estimated as
$\tau_{\rm c}\approx 3R_{\rm sh}/v_{\rm w} = 
10^4rR_{12}/v_3(r)$ s, where the radius of the WR star, $R_{\rm WR} = 
10^{12}R_{12}$ cm. The maximum energies of 
nuclei are determined by the condition $\tau_{\rm c} = \tau_{\rm acc}$. 
Therefore, we expect the cut-off in the power law spectrum of nuclei at energies
corresponding to
%
%
%
\begin{eqnarray}
\gamma_{\rm max} = 
{{3eZB_{\rm sh}R_{\rm sh}v_{\rm w}}\over{cAm_{\rm p}}}
\approx 5.3\times 10^3R_{12}rv_3(r)B_{\rm sh}(r).
\label{eq3}
\end{eqnarray}
\noindent
where $B_{\rm sh}$ is given by Eq.~(2).

The WR stars are the late stage of evolution of supermassive stars which have 
already lost their hydrogen envelopes. Therefore, their winds are mainly composed
of nuclei heavier than hydrogen. The heavy nuclei accelerated to Lorentz factors 
described by Eq.~\ref{eq1} and Eq.~\ref{eq3} lose nucleons due to the 
photo-disintegration 
process in collisions with the thermal photons from the massive stars if the energies 
of photons in the reference frame of the nuclei, 
$E_\gamma = 3k_{\rm B}T\gamma_A(1+\cos\theta) = 25T_5\gamma_6(1+\cos\theta)$ MeV, 
are above $\sim 2$ MeV, where $k_{\rm B}$ is the Boltzmann constant, 
$\theta$ is the angle between photon and nucleus. 
The above condition gives the lower limit on the Lorentz factor of accelerated 
nuclei, 
\begin{eqnarray}
\gamma_{\rm min}\approx 8\times 10^4/T_5(1+\cos\theta). 
\label{eq5}
\end{eqnarray}
Most efficient disintegrations occur when photon energies correspond to the 
maximum in the photo-disintegration cross-section of
nuclei, which is  at $\sim 20-30$ MeV with the half width of $\sim 10$ MeV
for nuclei with the mass numbers between helium and oxygen (main composition of 
the surface of the WR stars) .  
Let us estimate the efficiency of the photo-disintegration process of nuclei 
in the considered here scenario.
The average density of thermal photons from both stars (with similar temperatures), 
at the location of the shock, can be approximated by
\begin{eqnarray}
n_{\rm WR} \approx 
{{4\sigma_{\rm SB}T^3}\over{3ck_{\rm B}}}\left({{R_{\rm WR}}\over{R_{\rm sh}}}\right)^2 
       \approx 2\times 10^{16}{{T_5^3}\over{r^2}}~~{\rm ph.~cm}^{-3}.
\label{eq6}
\end{eqnarray}
where $\sigma_{\rm SB}$ is the Stefan-Boltzmann constant.
Then, the mean free path for dissociation of a single nucleon from a nucleus is
$\lambda_{A\gamma} = (n_{\rm WR}\sigma_{A\gamma})^{-1}\approx 
3.5\times 10^{10}r^2/(AT_5^3)~~{\rm cm}$, where for the photo-disintegration cross 
section the peak in the giant resonance is applied,  
$\sigma_{\rm A\gamma} = 1.45\times 10^{-27}A$ cm$^2$ (Karaku\l a \& Tkaczyk~1993).
For the considered here binary system WR 20a,
for which $R_{\rm sh} = a/2$ and so $r = 1.3$,  
the characteristic photo-disintegration time scale of nuclei with the mass number $A$, 
$\tau_{\rm A\gamma} = \lambda_{\rm A\gamma}/c\approx 30/A$ s, is much 
smaller than the acceleration time and the convection escape time 
of nuclei from the acceleration region at the shock (estimated above).
Nuclei with the initial mass numbers between 
4 (helium) and 16 (oxygen), accelerated to energies given by Eqs.~\ref{eq1}
and ~\ref{eq3}, should suffer complete fragmentation.
Note that nuclei, and protons from their disintegration, can not escape far away 
from the shock region. For typical values of the binary system WR 20a, 
their Larmor radii, $R_{\rm L}$, are significantly shorter
than the characteristic distance scale of the considered picture
defined by the semimajor axis of the binary system. 
Significant fraction of neutrons from disintegration of nuclei move toward
the surface of the massive stars since the probability of dissociation of a single nucleon
is the highest for the head on collisions of nuclei with thermal photons.  
These neutrons propagate along the straight lines and interact with the matter of  
stellar atmospheres. On the other hand, protons from disintegration of nuclei
are convected outside the binary system along the surface of the shock structure
in a relatively dense stellar winds.

The relativistic nuclei accelerated at the shock take a fraction, $\xi$, of the 
kinetic power of the two stellar winds,
\begin{eqnarray}
P_{\rm A} = \xi\dot{M}v_{\rm w}^2\approx 
6\times 10^{37}\xi \dot{M}_{-5}v_3^2(r)~~{\rm erg~s}^{-1}. 
\label{eq8}
\end{eqnarray}
The parameter $\xi$ is
determined by the efficiency of acceleration of nuclei and by the solid angle
$\Delta \Omega$ subtended by the \textit{active} part of the  shock (see Fig.~1). 
The {\it active} part of the shock is determined by the power of the stellar wind 
which falls onto the shock region in which nuclei can be accelerated above 
$\gamma_{\rm min}$ (see Eq.~\ref{eq5}).  
$\Delta \Omega$ can be estimated from 
the condition of acceleration of nuclei above $\gamma_{\rm min}$ 
by applying Eqs.~\ref{eq1} (for model I) or~\ref{eq3} (for model II) and
Eq.~\ref{eq2}. By comparing  Eq.~\ref{eq1} with 
Eq.~\ref{eq5} (for $\theta = 0^{\rm o}$), we estimate 
the maximum distance of the \textit{active} shock from the star on 
$R_{\rm max}\approx 4(B_3^2T_5L_{12}v_3^{1/2}(r))^{1/3}R_{\rm WR}$.
Similar procedure, comparison of Eq.~\ref{eq3} with 
Eq.~\ref{eq5}, allows us to estimate $R_{\rm sh}^{\rm max}$ in model II,
$R_{\rm max}\approx 10^2(R_{12}^2B_3T_5v_3(r)) R_{\rm WR}$.
A part of the shock solid angle at which nuclei are accelerated 
above $\gamma_{\rm min}$ is then $\Delta\Omega = 
2\pi(1-R_{\rm o}/R_{\rm max})/4\pi$, where
the distance of the shock in the plane of the binary system is approximated by 
$R_{\rm o}\approx a/2$.
For the considered above parameters of WR 20a, $B_3 = 1$, $T_5 = 0.4$, 
$v_3^\infty = 1$, $L_{12} = 0.3a/(10^{12} cm)$, $a/2 = 1.3R_{\rm WR}$, and 
$R_{\rm WR} = 20R_\odot$, $\Delta\Omega\sim 0.2$ and $0.5$ 
for the models I and II, respectively. 

Provided that complete disintegration of nuclei occurs, as envisaged above 
in this scenario, the flux of nucleons dissolved from nuclei in the case of 
monoenergetic acceleration in the reconnection regions (model I) is 
%
%
%
\begin{eqnarray}
N_{\rm n} = P_{\rm A}/m_{\rm p}\gamma_{\rm A}
\approx 4\times 10^{34}\xi\dot{M}_{-5}v_3^2(r)/\gamma_6~~{\rm N~s}^{-1}, 
\label{eq9}
\end{eqnarray}
In the case of nuclei with the power law spectrum (model II) the differential 
spectrum of nucleons is,
\begin{eqnarray}
dN_{\rm n}/d\gamma_{\rm n} dt = K\gamma_{\rm n}^{-\delta},
\label{eq9b}
\end{eqnarray}
\noindent
in the range $\gamma_2 = \gamma_{\rm max}$ (Eq.~\ref{eq3}) and applying 
$\gamma_1 = 10^1$, $\delta$ is the spectral index,
and the normalization constant is equal to 
$K = A P_{\rm A}/(Am_{\rm p}\ln(\gamma_2/\gamma_1))$
for $\delta = 2$ and  
$K = A P_{\rm A}(\delta - 2)/(Am_{\rm p}(\gamma_1^{2-\delta)}-
\gamma_2^{2-\delta}))$ for $\delta > 2$.
The spectral index of particles accelerated at the shock depends on the 
Alfvenic Mach number of the plasma flowing through the shock. It is defined as 
$M = v_{\rm w}/v_{\rm A}$ (see e.g. Schlickeiser~2002), 
where $v_{\rm A}$ is the Alfven velocity defined above.
For the parameters of WR 20a, we estimate the Alfvenic Mach number 
as a function of distance from the star on $M\approx r$, 
where the density of the wind and the magnetic field strength 
at the shock location are calculated above.
We conclude that the outer part of the shock, $r\gg 1$, 
fulfills the condition of the strong shock, $M\gg 1$, although in the 
inner part, the Alfven shock is probably created. Therefore, we consider the
power law spectrum of accelerated nuclei with indexes between 2 (characteristic 
for a strong shock) and 2.3 (for a weaker shock).

\begin{figure*}
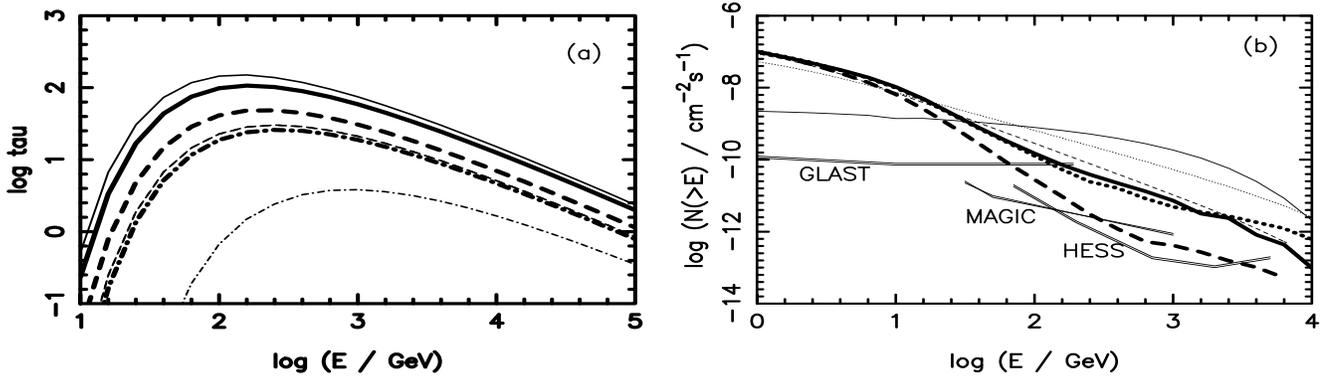

\vskip 5truecm
\includegraphics{fig2a.eps}
\includegraphics{fig2b.eps}
\caption{(a) The optical depths for $\gamma$-rays on $e^\pm$ pair production
in collision with stellar photons (as a function of their energy)
injected at two distances from the massive star WR 20a and for selected angles
measured from the direction defined by the centers of the stars, for 
$R = 1.3R_{\rm WR}$: $\alpha = 49^{\rm o}$ (thick full curve),
$70^{\rm o}$ (thick dashed), $90^{\rm o}$ (thick dot-dashed), and $3R_{\rm WR}$: 
$\alpha = 20^{\rm o}$ (thin full),
$85^{\rm o}$ (thin dashed), $150^{\rm o}$ (thin dot-dashed).
(b) The integral cascade $\gamma$-ray spectra escaping from the binary system 
(marked by the thick curves)
averaged over the observation angles produced by the primary $\gamma$-rays 
(marked by the thin curves) which in turn are injected by 
protons in collisions with the matter of the stellar winds via decay of pions. 
Protons are coming from dissociation of nuclei with the monoenergetic
spectrum (thick full curve) and the power law spectra with the spectral indexes equal 
to 2. (thick dotted curve) and 2.3 (thick dashed curve). The cascade $\gamma$-ray 
spectra are normalized to the GeV flux observed from the EGRET sources (GeV J1025-5809
and GeV J1047-5840, Lamb \& MaComb~1997) in the direction of WR 20a
The double thin curves show the sensitivity limits for the GLAST (1 year observation), 
MAGIC, and HESS telescopes ($5\sigma$ 50 hrs observation, $> 10$ events), 
e.g. Lorenz~2001.}
\label{fig2}
\end{figure*}

Since the winds of the massive stars are very dense, protons from
disintegration of nuclei have chance to interact with the matter of the winds. 
The characteristic time scale for 
collision of relativistic protons with the matter of the wind is 
$\tau_{\rm pp}\approx (c\sigma_{\rm pp}\rho)^{-1}\approx 
10^4 r^2v_3(r)/\dot{M}_{-5}$ s,
where $\rho$ is the density of the stellar wind estimated above.
For the applied above parameters of the wind, $v_3^\infty = 1$ and $\dot{M}_{-5} = 3$,
$\tau_{\rm pp}$ is shorter than escape time scale of protons along the shock,
$\tau_{\rm c}$, at distances less than $r\approx 5$, i.e. in the main part of the 
\textit{active} shock. 

We calculate the $\gamma$-ray spectra from decay of pions which are produced 
by protons in collisions with the matter of the wind by integrating the injection
spectra of protons (monoenergetic or the power law) over the \textit{active} part of 
the shock and applying  
the scale break model for hadronic interactions developed 
by Wdowczyk \& Wolfendale (1987) which is suitable for the considered energies
of relativistic protons. Only single interaction of proton with the matter has been 
included.
These high energy $\gamma$-rays originate relatively close to the surface of the massive 
stars, $r\sim 1.3-5$, and therefore they can suffer absorption in the thermal radiation 
coming from the stellar surfaces. In Fig.~2a, we show the optical depths for 
$\gamma$-rays as a function of their energies. It is clear that most of the 
$\gamma$-rays with energies
above a few tens GeV are absorbed and initiate the inverse Compton (IC) $e^\pm$ pair 
cascades in the radiation of the massive stars.
In order to determine the final $\gamma$-ray spectra which escape from the binary 
system, we apply the Monte Carlo code developed by Bednarek (2000) which follow
the IC $e^\pm$ pair cascade in the anisotropic radiation field of the massive star.
It is assumed that primary $\gamma$-rays are produced by protons isotropically
at the \textit{active} part of the shock.
The phase averaged $\gamma$-ray spectra escaping from the binary system are shown 
in Fig.~2b for the monoenergetic and the power law spectra of injected protons,
assuming the distance to the WR 20a equal to 5 kpc (the average 
value estimated from the reported range (2.5-8) kpc, Churchwell et al.~2004).
These spectra have been normalized to the $\gamma$-ray fluxes observed from
the EGRET sources, 2EG J1021-5835 and GeV J1025-5809
and 2EG J1049-5847 and GeV J1047-5840 (see Thompson et al.~1995, Lamb \& MaComb~1997),
which are equal to $\sim 10^{-7}$ ph cm$^{-2}$ s$^{-1}$ above 1 GeV.
Based on these normalizations we derive the required acceleration efficiencies of
nuclei at the shock equal to $\xi\sim 13.5\%$ (for the monoenergetic injection)
and $\sim 12.3\%$ and $\sim 8\%$ for the power law injection with the spectral
indexes 2 and 2.3, respectively. The $\gamma$-ray spectra predicted by this model
(shown in Fig.~2b) should be detectable above 100 GeV by the Cherenkov telescopes on 
the Southern hemisphere such as the HESS and CANGAROO III or at lower energies by 
the MAGIC type telescope (e.g. HESS II).

The $\gamma$-ray fluxes produced by protons at the shock region in collisions
with the matter of the winds via decay of pions are also accompanied by the high 
energy neutrinos.
Let us estimate the muon neutrino event rate from our example binary system WR 20a. 
The spectra of muon neutrinos, calculated exactly as the primary spectra of 
$\gamma$-rays, are shown in Fig.~3a. They are clearly above the atmospheric neutrino 
background (ANB) and above the sensitivity limit of the 1 km$^2$ neutrino detector
of the IceCube type. In the case of monoenergetic injection of nuclei they are 
also above the present sensitivity of the AMANDA II neutrino detector. 
However due to the localization of the source on the Southern hemisphere
this model cannot be at present excluded.
The number of events produced by the muon neutrinos in the 1 km$^2$ neutrino 
detector of the IceCube type  can be estimated from 
\begin{eqnarray}
N_{\mu} = {{S}\over{4\pi D^2}} \int 
P_{\nu\rightarrow\mu}(E_\nu) {{dN_\nu}\over{dE_\nu dt}} dE_\nu,
\label{eq12a}
\end{eqnarray}
where $S = 1$ km$^2$ is the surface of the detector,
$P_{\nu\rightarrow\mu}(E_\nu)$ is the energy dependent detection 
probability of muon neutrino (Gaisser \& Grillo~1987).

Applying estimated above values for acceleration efficiency, the distance to WR 20a 
equal to 5 kpc, and typical wind parameters of the WR stars, $\dot{M}_{-5} = 3$, 
$v_3^\infty = 1$, we 
predict in the case of the model I $N_\mu\sim 23.5~(23)$ muon neutrino events inside 
1 km$^2$ detector in 1 yr for the case of neutrinos arriving
from the directions close to the horizon (no absorption inside the Earth) and from 
the nadir (partial absorption inside the Earth). 
The neutrino event rate predicted in the case of model II is   
$N_\mu\sim 21.6~(20.5)$ and  $\sim 3.7~(3.5)$ muon neutrino events in 1 km$^2$ in 
1 yr provided that nuclei are accelerated with the power law spectrum and spectral 
indexes equal to 2 and 2.3, respectively. In these calculations we applied the Earth
shadowing factors from Gandhi~(2000). The high event rates predicted for the 
nuclei injected with the flat spectra should be easily tested by the future 
0.1 km$^2$ Antares telescope.

\begin{figure}
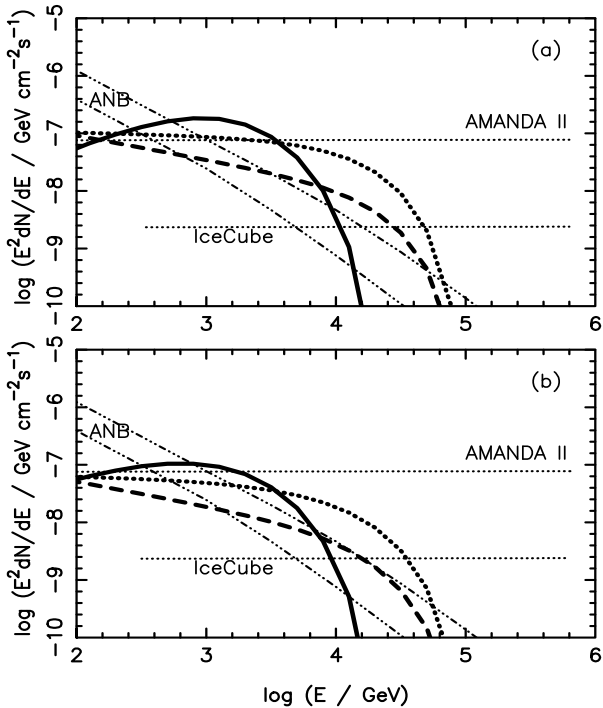

\vskip 9.5truecm
\includegraphics{fig3a.eps}
\includegraphics{fig3b.eps}
\caption{The differential spectra of muon neutrinos produced by protons interacting 
with the matter of the stellar winds (a) and by neutrons interacting with the 
matter of the stellar atmosphere (b). Specific neutrino spectra are 
calculated for the parameters as the corresponding $\gamma$-ray spectra shown 
in Fig.~2b (these same curve styles). The dot-dot-dot-dashed curves 
indicate the atmospheric neutrino background 
(horizontal - upper curve, and vertical - lower curve)
within a $1^o$ of the source (Lipari 1993),
the thin dotted line shows the 3 yr sensitivity of the IceCube detector (Hill 2001),
and the thick dotted line shows the upper limit obtained by AMANDA II detector 
(Ackermann et al. 2004).}
\label{fig3}
\end{figure}

A fraction of neutrons, $\eta$, dissolved from nuclei 
can fall onto the surfaces of the massive stars. 
$\eta$ can be calculated as a function of the shock distance from the WR star, 
$r$, and the Lorentz factor of neutrons, $\gamma_{\rm n}$, from the formula

\begin{eqnarray}
\eta (\gamma_{\rm n}) = {{2F(\cos\theta_{\rm D},1)}\over{F(\cos\theta_{\rm D},1) + 
F(\cos\theta_{\rm D}, - \cos\theta_{\rm D})}}
\label{eq13}
\end{eqnarray}

\noindent
where $F(x,y) = \int_{\rm x}^{\rm y}\sigma_{\rm A\gamma} n_{\rm WR}
(\varepsilon, \cos\theta) (1 + \cos\theta) d\varepsilon d\cos\theta$, $\theta_{\rm D}$ is 
the angle intercepted by the massive star observed from the distance $r$.
We have performed such calculations and found that $\eta$ weekly depends on the 
Lorentz factor of nuclei, except for the Lorentz factors for which efficient 
fragmentation starts to occur. It is equal to
$\eta\sim 0.52$, 0.33, 0.22 for $R_{\rm sh} = 1.3R_{\rm WR}$, $2R_{\rm WR}$, 
and $3R_{\rm WR}$, respectively. 
The Lorentz factors of neutrons extracted from nuclei are equal to the Lorentz 
factors of the parent nuclei $\gamma_{\rm A} = \gamma_{\rm n}$.
Neutrons which fall onto the surface of the massive stars produce
charged pions in collisions with the matter of the stellar atmospheres.
Pions from the first interaction are produced at the characteristic densities 
$\rho \sim 3\times 10^{14}$ cm$^{-3}$ (estimated from the p-p cross section
and applying the characteristic dimension of the stellar atmosphere 
$\sim 0.1 R_{\rm WR}$). 
In such conditions pions with the Lorentz factors $\gamma_{\rm n} = 10^6$ decay before 
interacting  with the matter since they pass only $\sim 0.3$ g cm$^{-2}$.
This conclusion has been checked by performing the Monte Carlo simulations 
of the interaction process of neutrons with the stellar atmosphere
which preliminary results has been published in Bartosik, Bednarek \& 
Sierpowska~(2003).

We have calculated the spectra of neutrinos produced by neutrons in the atmospheres 
of the massive stars applying the scale break model for hadronic interactions, and
taking into account the fraction of neutrons falling onto the stars $\eta$,
and the multiple interactions of neutrons with the matter of the stellar atmospheres
(see Fig.~3b).
The neutrino event rates, for the acceleration models considered above, are 
estimated on $N_\mu\sim 12.8~(12.6)$ events for the monoenergetic spectrum of neutrons, 
and $11~(10.4)$ and $2.9~(2.7)$ events for the spectra of neutrons which are close 
to the power law spectra of their parent nuclei with 
the index 2 and 2.3, respectively.
These event rates are lower than that produced by protons since only a fraction of 
neutrons interact. From another site this effect is partially compensated by the 
multiple interactions of neutrons with the matter of the stellar atmospheres.
Note moreover that neutrinos produced by neutrons are emitted within the limited 
range of angles, $\alpha$, 
around the plane of the binary system defined by, $\tan\alpha\sim 
\sqrt{R_{\rm max}^2-R_{\rm o}^2}/(R_{\rm o}-R_{\rm WR})$ 
(see values defined above), i.e within $\alpha\sim 79^{\rm o}$ for 
$R_{\rm max} = 2R_{\rm WR}$.
This angle is much larger than the eclipsing angle of the WR 20a binary system
$\sim 40^{\rm o}$. Therefore neutrinos produced by neutrons 
can be also potentially observed from non-eclipsing binaries (see  
Fig.~\ref{fig1}). However, in the case of WR 20a, which inclination angle is
$\sim 74.5^{\rm o}$ (Bonanos et al.~2004), the observer is located inside 
the eclipsing cone. 
The neutrino signal produced by neutrons should be modulated with the 
orbital period of the binary system as postulated by the emission angle $\alpha$. 
This may help extracting the neutrino signal produced by neutrons from the 
atmospheric background.

The $\gamma$-ray and neutrino signals from such close binaries as 
WR 20a can be strongly variable due to the very unstable 
position of the wind collision region (short cooling time scales of 
thermal plasma). Small change in one of the winds may push the wind
collision region onto the surface of one of the stars changing somewhat
conditions in the acceleration region.

\section*{Acknowledgments}
I would like to thank the anonymous referee and V. Bosch-Ramon for many useful comments.
This work is supported by the MNiI grant No. 1P03D 010 28 and the KBN grant 
PBZ KBN 054/P03/2001. 


\end{document}